\documentclass[final,1p,times]{elsarticle} 

\usepackage{graphicx}
\usepackage{amssymb} 
\usepackage{amsthm} 
\usepackage{lineno}
\usepackage[figuresright]{rotating}
\usepackage{wrapfig}
\usepackage{epsfig}
\usepackage{dcolumn}
\usepackage{bm}
\usepackage{color}

\newcommand{\be}{\begin{equation}}
\newcommand{\ee}{\end{equation}}
\newcommand{\bea}{\begin{eqnarray}}
\newcommand{\eea}{\end{eqnarray}}

\newcommand{\bk}{\mathbf{k}}
\newcommand{\bq}{\mathbf{q}}

\journal{Nuclear Physics A} 

\begin{document}

\begin{frontmatter} 

\title{A running coupling explanation\\ of the surprising transparency of the QGP at LHC}
\author{Alessandro Buzzatti}
\ead{buzzatti@phys.columbia.edu}
\author{Miklos Gyulassy}
\address{Department of Physics, Columbia University, 538 West 120th Street, New York, NY 10027, USA}

\begin{abstract} 
The CUJET Monte Carlo Jet Energy loss model is applied to predict the jet flavor, centrality and density dependence of the nuclear modification factor $R_{AA}$ at RHIC and LHC. Running coupling effects due to combined $x$, $k_\perp$ and $q_\perp$ evolution are included for the first time in the dynamical DGLV opacity expansion framework and are shown to provide a natural dynamical QCD tomographic solution to the surprising transparency of the quark gluon plasma produced at LHC as suggested by $p_\perp>10$ GeV $R_{AA}$ data from ALICE, ATLAS, and CMS.
\end{abstract} 

\end{frontmatter}

\section{Introduction}

In this proceeding we report results of a recent update of the CUJET \cite{CUJET} pQCD tomographic model, where running coupling effects have been taken into account.
We remind that CUJET is a model developed as part of the ongoing Department of Energy JET Topical Collaboration \cite{JETColl} effort to construct more powerful numerical codes, which are
necessary to reduce the large theoretical and numerical systematic uncertainties that have hindered so far quantitative jet tomography.
CUJET extends the development of the the GLV, DGLV and WHDG \cite{GLV,DGLV,WHDG} opacity series approaches by including several dynamical features that allow the evaluation systematic theoretical uncertainties such as sensitivity to formation and decoupling phases of the QGP evolution, local running coupling and screening scale variations, and other effects out of reach with analytic approximations.

The model is here applied to predict the nuclear modification factor $R_{AA\rightarrow a\rightarrow f}(y\approx 0, p_T, \sqrt{s} ,{\cal C}=0-5\%)$, for a variety of jet parton flavors $a=g,u,c,b$ and final fragments $f=\pi,D,B$, over a broad $p_T$ kinematic range at mid-rapidity for central collisions at $\sqrt{s}=0.2,2.76$ GeV as observed at RHIC \cite{RHIC} and LHC \cite{LHC}.
By implementing running coupling effects, justified by the broad range of energies at play and already suggested in \cite{Betz}, we want to show that this relatively simple physical phenomenon could account for the surprising transparency of the QGP at LHC, as recently observed by the ALICE and CMS experiments \cite{LHClatest}.

\section{CUJET model}
The Monte Carlo techniques implemented in CUJET allow the model to compute finite order in opacity $n>1$ contributions to the jet medium induced gluon radiative spectrum.
As in previous works \cite{CUJET}, however, we limit our studies to the first order in opacity; furthermore, we model the interaction potential between the jet and the medium with the pure HTL dynamical model \cite{MD}.
Neither of these approximations are shown to qualitatively alter the results, once rescaling of the effective coupling constant---or $\alpha_0$ as we will see shortly---is taken into account.
Both radiative and elastic contributions to the energy loss are considered, and fluctuations of the radiated gluon number included via Poisson expansion (incoherent emission).

The plasma is assumed inhomogeneous (Glauber profile) and nonstatic (1+1D Bjorken expansion), and its density profile is constrained solely by the initial observed rapidity density $dN/dy$, equal here to $1000$ for $\sqrt{s}=0.2$ GeV RHIC collisions and $2200$ for $\sqrt{s}=2.76$ GeV LHC collisions; we also set the initial thermalization time $\tau_0=1$ fm/c and take $T\approx 100$ MeV as the characteristic temperature at which the jet decouples from the medium.

\section{Running coupling}
As we already mentioned, recent preliminary results published by the ALICE and CMS collaborations \cite{LHClatest} appear to indicate a steeper rise of $R_{AA}$ in the range of momenta $30-100$ GeV.
In addition to this, LHC extrapolations performed using an effective static coupling constant fit to RHIC results seem to systematically overpredict the quenching suffered by the jets in the plasma \cite{WHMG}.
These findings motivated us to relax the effective fixed alpha approximation and utilize a one-loop order running coupling, parametrized as follows \cite{Zakharov_RunningCoupling}:
\be
\alpha_s(Q^2)= \left\{ \begin{matrix}
\alpha_0\equiv\frac{2\pi}{9\ln(Q_0/\Lambda_{QCD})} \; (Q\leq Q_0) \; ;\\
\frac{2\pi}{9\ln(Q/\Lambda_{QCD})} \;(Q>Q_0) \; .\\
\end{matrix}\right.
\ee
Here we choose to keep $\alpha_0$ as the only free parameter of the model.
The choice of scale $Q$, of the order of $1$ GeV, is somewhat arbitrary: of the three powers of $\alpha_s$ that occur in the DGLV integral, two are related to the medium scattering vertex and one to the radiated gluon. We let the first two to scale as $\bq_\perp^2$, the transverse momentum exchanged between the jet and the medium, and the latter to scale approximately with $\hat{t}\approx\bk_\perp^2 / (x(1-x))$, where $\bk$ represent the transverse momentum of the radiated gluon and $x$ is its fraction of plus-momentum.
In order to account for this source of systematic uncertainty, we let the scale vary of $\pm 25\%$, while fixing the parameter $\alpha_0$ to fit one chosen pion $R_{AA}^{LHC}(p_T=40GeV)=0.35$ point.

At the same time, we also include running coupling effects in the elastic contribution to the energy loss \cite{PeignePeshier}. The two powers of $\alpha_s$, in this case, scale as $ET$ and $\mu^2$, with $E$,$T$,$\mu$ the energy of the jet, the temperature of the plasma and the Debye screening mass respectively.

Of all these contributions, the most relevant comes from the radiated gluon vertex, $\alpha_s(\bk_\perp^2 / (x(1-x)))$; in particular, this helps explaining the faster rise in the shape of pion $R_{AA}$.
In fact, the hard tails of the opacity expansion spectrum make the $\bk_\perp$ integral, whose upper kinematic limits are a function of the energy of the jet $E$, more sensitive to the high $\bk_\perp$ region where $\alpha_s$ is smaller.
This, in turn, clearly modifies the energy dependence of the nuclear modification factor by reducing the relative energy loss at large energies; as a consequence, $R_{AA}$ will show a steeper rise with $p_\perp$ than in the fixed coupling scenario.

\section{Results}

\begin{figure*}[tbh]
\centering
\includegraphics[width=2.8in
]{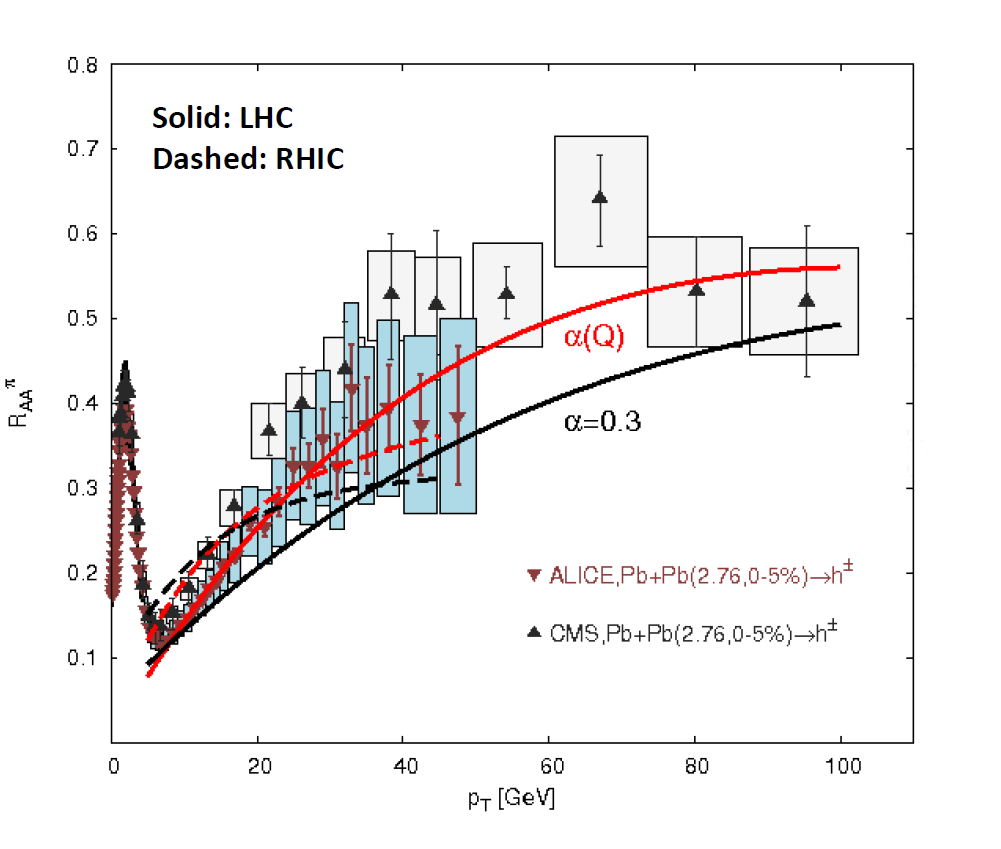}
\hspace{0.25in}
\includegraphics[width=1.8in
]{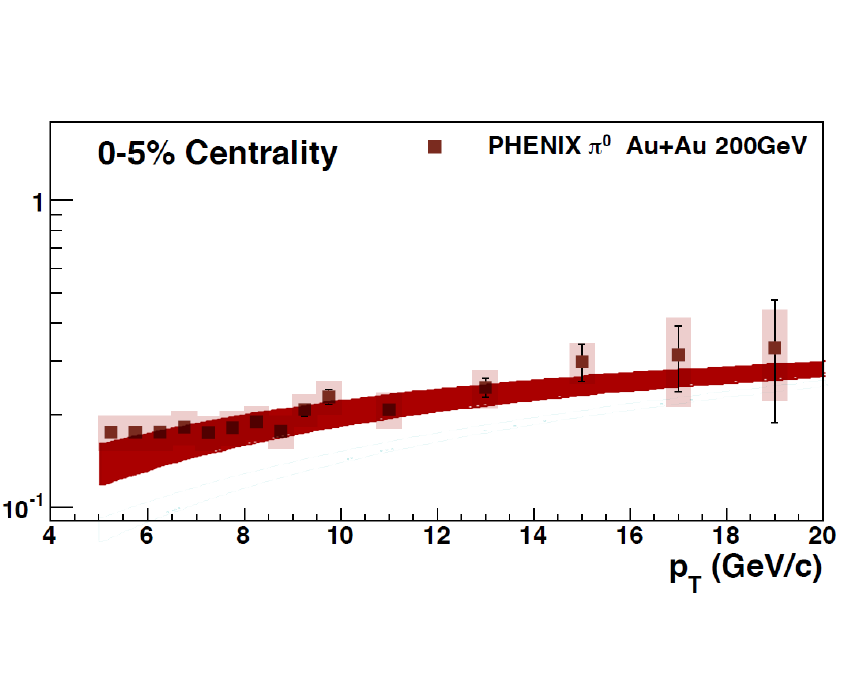}
\caption{(Left) Pion $R_{AA}$ at LHC (solid lines) and RHIC (dashed lines). In black the fixed coupling CUJET results, constrained at $p_T=10$ GeV RHIC with $\alpha_s=0.3$. In red the running coupling CUJET results, constrained at $p_T=40$ GeV LHC with $\alpha_0=0.4$. Central $0\%-5\%$ preliminary ALICE and CMS $h\pm$
LHC data \cite{LHClatest} (brown and gray triangles, respectively) are compared to predictions. (Right) Notice the accordance with data \cite{PHENIX} of the RHIC Pion $R_{AA}$ running coupling result in the range of energies $5-20$ GeV.}
\label{Fig1}
\end{figure*}

In CUJET, the parameter $\alpha_0$ in the expression for the running coupling is the only free parameter that we fit to the data: constraining one reference $p_T=40$ GeV point of pion
$R_{AA}^\pi=0.35$ at LHC, we set $\alpha_0=0.4$. The backward extrapolation to RHIC is then parameter free, assuming that $\alpha_0$ does not vary with $\sqrt{s}$. This is an inversion of the standard practice to fit the data at RHIC and extrapolate to LHC, however is needed since the physics we now want to probe spans over a much broader range of energies, inaccessible at RHIC.
The results are shown in Fig.~1.
Observing the figure on the left, it is evident that the overall shape of $R_{AA}$ across the broad range of $p_T$ under consideration is changed with respect to the previous fixed coupling results. Besides appreciating the more satisfactory agreement with the data, both at LHC and RHIC (in the latter case our predictions are almost left unchanged given the restricted range of energies at play, as shown in the panel on the right), it is surprising to note how the effective energy dependence itself of the energy loss appears to be modified.

\begin{figure*}[tbh]
\centering
\includegraphics[width=2.42in
]{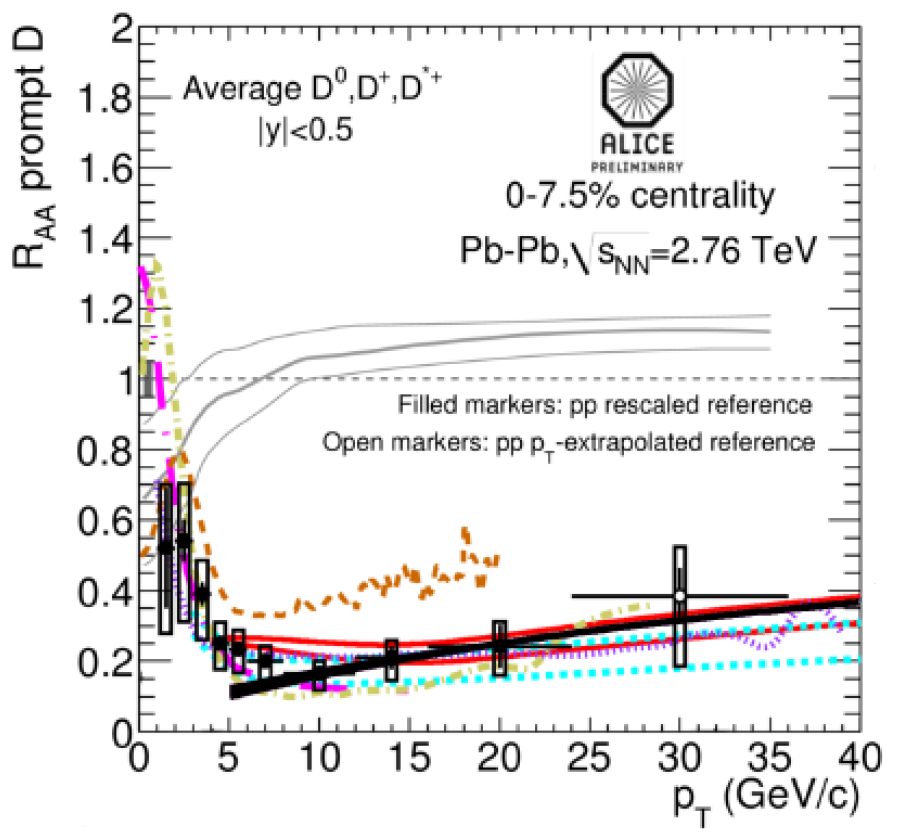}
\hspace{0.25in}
\includegraphics[width=2.18in
]{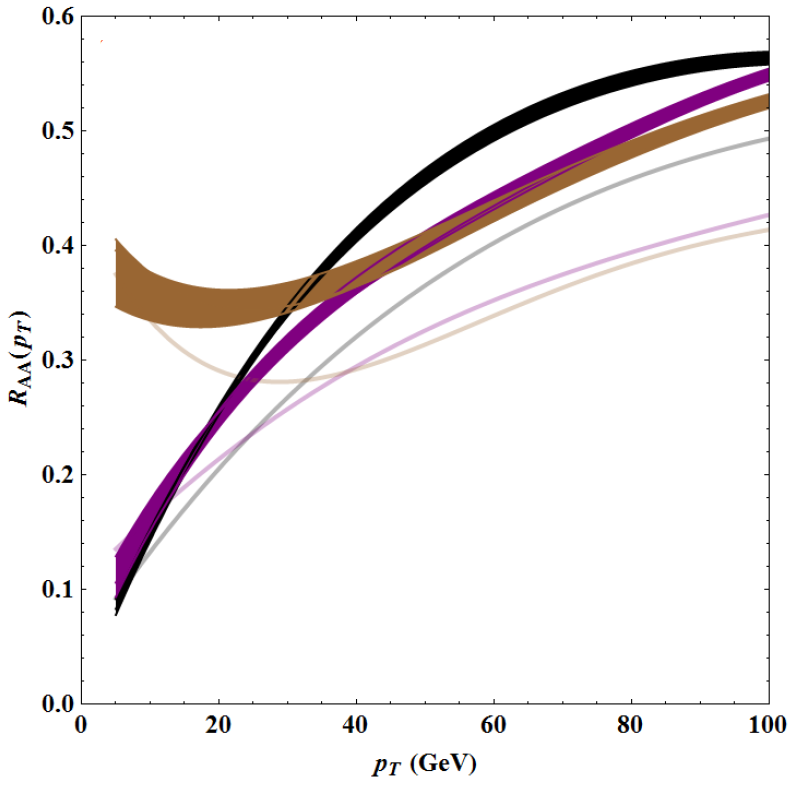}
\caption{(Left) D meson $R_{AA}$ at LHC. In black the running coupling CUJET results, constrained by the Pion fit shown in Fig.~1. Central $0\%-7.5\%$ preliminary ALICE LHC data \cite{LHClatest} are compared to predictions. (Right) Illustration of jet flavor tomography level crossing pattern of central $R_{AA}$ versus $p_T$ for Pions, D and B mesons. LHC Pb+Pb predictions are shown in solid color, RHIC Au+Au results are shown in faded colors. The opacity is constrained at LHC, given
  $dN/dy(LHC)=2200$, by a fit to a reference
  point $R_{AuAu}^\pi(p_T=40\;{\rm GeV})=0.35$ setting $\alpha_0=0.4$.}
	\label{Fig2}
\end{figure*}

In Fig.~2, we are showing the flavor dependence of $R_{AA}$ and in particular the comparison with the preliminary D meson data from ALICE \cite{LHClatest}. It is worth restating that the heavy meson predictions plotted on the left are constrained by the same parameters already fit to the pion results, a fact that seems to indicate a remarkable robustness of CUJET, at least when compared to the presently available data.
On the right panel, we can instead observe the full flavor dependence of the nuclear modification factor: the most striking feature is the inversion of the $\pi<D<B$ $R_{AA}$ hierarchy ordering at high $p_T$, already seen in the case of the fixed coupling constant in \cite{CUJET}. This effect is mostly due to the steeper initial invariant jet distributions of c and b jets at RHIC \cite{Vogt}.

\section{Conclusions}

The CUJET model has been applied here to study the flavor and $\sqrt{s}$ dependence of the nuclear modification factors for central collisions at mid-rapidity. The inclusion of running coupling effects in the model has vastly improved the agreement with the recent LHC data without affecting the RHIC sector in the range of energies currently probed. With one free parameter ($\alpha_0$) used to fit the pion data at LHC, we have predicted the same level crossing pattern of $R_{AA}$ already seen in the case of fixed coupling. At the same time, we have shown that the theoretical uncertainties relative to the choice of running scales do not affect qualitatively the results.
Further data, especially for heavy mesons (both at RHIC and LHC) and extended energy range (RHIC) may give additional indication of running effects and lead to stronger conclusions with respect to the energy dependence of the energy loss itself.

{Acknowledgments}: We acknowledge support by US-DOE Nuclear Science Grant No.
DE-FG02-93ER40764. 

\bibliographystyle{elsarticle-num}

\begin{thebibliography}{00}

\bibitem{CUJET}
  A.~Buzzatti and M.~Gyulassy, Phys. Rev. Lett. 108, 022301 (2012);
	A.~Buzzatti and M.~Gyulassy, Hard Probes 2012 proceeding, to be published;
  A.~Buzzatti and M.~Gyulassy, Nucl. Phys. A 855, 307 (2011);
  to be published.

\bibitem{JETColl}
  (JET) Topical Collaboration on Jet and Electromagnetic Tomography,
  http://www-nsdth.lbl.gov/jet/.

\bibitem{GLV}
  (GLV) M.~Gyulassy, P.~Levai and I.~Vitev,
  Nucl. Phys. B 594, 371 (2001).

\bibitem{DGLV}
  (DGLV) M.~Djordjevic and M.~Gyulassy,
  Nucl. Phys. A 733, 265 (2004).

\bibitem{WHDG}
  (WHDG) S.~Wicks, W.~Horowitz, M.~Djordjevic and M.~Gyulassy,
  Nucl. Phys. A 784, 426 (2007).

\bibitem{RHIC}
  J.~Adams {\it et al.}  [STAR Collab.],
  Nucl. Phys. 757, 102 (2005);
  K.~Adcox {\it et al.}  [PHENIX Collab.],
  Nucl. Phys. A 757, 184 (2005).

\bibitem{LHC}
  [ALICE Collab.];
  [CMS Collab.].
		
\bibitem{Betz}
  B.~Betz and M.~Gyulassy,
  arXiv:arXiv:1201.0281 [nucl-th].

\bibitem{LHClatest}
  B. Abelev {\it et al.} [ALICE Collaboration],
  arXiv:1205.5761 [nucl-ex];
	K. Šafa?ík for the ALICE Collaboration,
	Quark Matter 2012;
  S. Chatrchyan {\it et al.} [CMS Collaboration],
  arXiv:1202.2554 [nucl-ex].

\bibitem{MD}
  M.~Djordjevic,
  Phys. Rev. C 80, 064909 (2009);
  M.~Djordjevic and U.~W.~Heinz,
  Phys. Rev. Lett. 101, 022302 (2008).

\bibitem{WHMG}
  W.~A.~Horowitz and M.~Gyulassy,
  arXiv:1104.4958 [hep-ph].
	
\bibitem{Zakharov_RunningCoupling}
  B.~G.~Zakharov,
  JETP Lett. 88, 781 (2008)
	
\bibitem{PeignePeshier}
  S.~Peigne and A.~Peshier,
  Phys. Rev. D 77, 113017 (2008)

\bibitem{PHENIX}
  A.~Adare {\it et al.}
  Phys. Rev. Lett.  101, 232301 (2008);
	T. Sakaguchi for the PHENIX Collaboration,
	Quark Matter 2012.

\bibitem{Vogt}
  M.~Cacciari, P.~Nason and R.~Vogt,
  Phys. Rev. Lett. 95, 122001 (2005);
  R.Vogt Private communication.



\end{thebibliography}

\end{document}